\pdfoutput=1
\documentclass[sigconf]{acmart}
\usepackage{booktabs} % For formal tables
\usepackage{subcaption}
\usepackage{url}

\acmDOI{10.475/123_4}

\acmISBN{123-4567-24-567/08/06}

\acmConference[C+J'19]{Computation+Journalism Symposium}{February 2019}{Miami, Florida USA}
\acmYear{2019}
\copyrightyear{2019}

\acmArticle{4}
\acmPrice{15.00}

\begin{document}
\title{Differences between Health Related News Articles from Reliable and Unreliable Media}

\author{Sameer Dhoju, Md Main Uddin Rony, Naeemul Hassan}

\affiliation{Department of Computer Science and Engineering, The University of Mississippi}

% \authornote{Dr.~Trovato insisted his name be first.}
% \affiliation{%

% \author{Second Author}
% \authornote{The secretary disavows any knowledge of this author's actions.}
% \affiliation{%

% \author{Third Author}
% \authornote{This author is the
%   one who did all the really hard work.}
% \affiliation{%

% \author{Fourth Author}

\renewcommand{\shortauthors}{S. Dhoju et al.}

\begin{abstract}
In this study, we examine a collection of health-related news articles published by reliable and unreliable media outlets. Our analysis shows that there are structural, topical, and semantic differences in the way reliable and unreliable media outlets conduct health journalism. We argue that the findings from this study will be useful for combating health disinformation problem.
\end{abstract}



% \keywords{ACM proceedings, \LaTeX, text tagging}

\maketitle

\section{Introduction}
\label{sec-introduction}
Of the $20$ most-shared articles on Facebook in $2016$ with the word ``cancer'' in the headline, more than half the reports were discredited by doctors and health authorities~\cite{independent}. The spread of health-related hoaxes is not new. However, the advent of Internet, social networking sites (SNS), and click-through-rate (CTR)-based pay policies have made it possible to create hoaxes/``fake news'', published in a larger scale and reach to a broader audience with a higher speed than ever~\cite{gigaom}. Misleading or erroneous health news can be dangerous as it can lead to a critical situation. ~\cite{houston2018measles} reported a measles outbreak in Europe due to lower immunization rate which experts believed was the result of anti-vaccination campaigns caused by a false news about MMR vaccine. Moreover, misinformation can spoil the credibility of the health-care providers and create a lack of trust in taking medicine, food, and vaccines. Recently, researchers have started to address the fake news problem in general~\cite{shu2017fake,lazer2018science}. However, health disinformation is a relatively unexplored area. According to a report from Pew Research Center~\cite{pew2014}, $72\%$ of adult internet users search online for information about a range of health issues. So, it is important to ensure that the health information which is available online is accurate and of good quality. There are some authoritative and reliable entities such as National Institutes of Health (NIH)~\footnote{https://www.nih.gov/} or \emph{Health On the Net}~\footnote{https://www.hon.ch/en/} which provide high-quality health information. Also, there are some fact-checking sites such as Snopes.com~\footnote{https://www.snopes.com/} and Quackwatch.org ~\footnote{http://www.quackwatch.org/} that regularly debunk health and medical related misinformation. Nonetheless, these sites are incapable of busting the deluge of health disinformation continuously produced by unreliable health information outlets (e.g., RealFarmacy.com, Health Nut News). Moreover, the bots in social networks significantly promote unsubstantiated health-related claims~\cite{usnewsbot}. Researchers have tried developing automated health hoax detection techniques but had limited success due to several reasons such as small training data size and lack of consciousness of users ~\cite{ghenai2017catching,kostkova2016vac,ghenai2018fake,vraga2017using}.

The objective of this paper is to identify discriminating features that can potentially separate a reliable health news from an unreliable health news by leveraging a large-scale dataset. We examine how reliable media and unreliable media outlets conduct health journalism. First, we prepare a large dataset of health-related news articles which were produced and published by a set of reliable media outlets and unreliable media outlets. Then, using a systematic content analysis, we identify the features which separate a reliable outlet sourced health article from an unreliable sourced one. These features incorporate the structural, topical, and semantic differences in health articles from these outlets. For instance, our structural analysis finds that the unreliable media outlets use clickbaity headlines in their health-related news significantly more than what reliable outlets do. Our semantic analysis shows that on average a health news from reliable media contains more reference quotes than an average unreliable sourced health news. We argue that these features can be critical in understanding health misinformation and designing systems to combat such disinformation. In the future, our goal is to develop a machine learning model using these features to distinguish unreliable media sourced health news from reliable articles.

% To accomplish this multi-facet analysis, we 
\section{Related Work}
\label{sec-related}
% \subsection{Health Journalism}
There has been extensive work on how scientific medical research outcomes should be disseminated to general people by following health journalism protocols ~\cite{kagawa2003strategy,shuchman1997medical,schwitzer2008us,dalmer2017questioning,de2016responsible}. For instance, \cite{lopes2009journalists} suggests that it is necessary to integrate journalism studies, strategic communication concepts, and health professional knowledge to successfully disseminate professional findings. Some researchers particularly focused on the spread of health misinformation in social media. For example, \cite{ghenai2017catching} analyzes Zika~\footnote{\url{https://en.wikipedia.org/wiki/Zika_virus}} related misinformation in Twitter. In particular, it shows that tracking health misinformation in social media is not trivial, and requires some expert supervision. It used crowdsource to annotate a collection of Tweets and used the annotated data to build a rumor classification model. One limitation of this work is that the used dataset is too small (6 rumors) to make a general conclusion. Moreover, it didn't consider the features in the actual news articles unlike us. \cite{ghenai2018fake} examines the individuals on social media that are posting questionable health-related information, and in particular promoting cancer treatments which have been shown to be ineffective. It develops a feature based supervised classification model to automatically identify users who are comparatively more susceptible to health misinformation. There are other works which focus on automatically identifying health misinformation. For example, \cite{kinsora2017creating} developed a classifier to detect misinformative posts in health forums. One limitation of this work is that the training data is only labeled by two individuals. Researchers have also worked on building tools that can help a user to easily consume health information. \cite{kostkova2016vac} developed the ``VAC Medi+board'', an interactive visualization platform integrating Twitter data and news coverage from a reliable source called MediSys\footnote{\url{http://medisys.newsbrief.eu}}. It covers public debate related to vaccines and helps users to easily browse health information on a certain vaccine-related topic.

Our study significantly differs from these already existing researches. Instead of depending on a small sample of health hoaxes like some of the existing works, we take a different approach and focus on the source outlets. This gives us the benefit of investigating with a larger dataset. We investigate the journalistic practice of reliable and unreliable health outlets, an area which has not been studied according to our knowledge.
\section{Data Preparation}
\label{sec-data}
For investigating how reliable media outlets and unreliable outlets portray health information, we need a reasonably sized collection of health-related news articles from these two sides. Unfortunately, there is not an available dataset which is of adequate size.
% or does contain rich contextual information such as graphical media or huperlinks. The contextual information are critical for identifying rich patterns beyond linguistic styles.
For this reason, we prepare a dataset of about $30,000$ health-related news articles disseminated by reliable or unreliable outlets within the years $2015-2018$. Below, we describe the preparation process in detail.

\subsection{Media Outlet Selection}
The first challenge is to identify reliable and unreliable outlets. The matter of reliability is subjective. We decided to consider the outlets which have been cross-checked as reliable or unreliable by credible sources.
\subsubsection{Reliable Media}
We identified $29$ reliable media outlets from three sources-- \textbf{i)} $11$ of them are certified by the Health On the Net~\cite{honcode}, a non-profit organization that promotes transparent and reliable health information online. It is officially related with the World Health Organization (WHO)~\cite{who}. \textbf{ii)} $8$ from U.S. government's health-related centers and institutions (e.g., CDC, NIH, NCBI), and \textbf{iii)} $10$ from the most circulated broadcast~\cite{broadcast} mainstream media outlets (e.g., CNN, NBC). Note, the mainstream outlets generally have a separate section for health information (e.g., \url{https://www.cnn.com/health}). As our goal is to collect health-related news, we restricted ourselves to their health portals only.
% We collected 38 Facebook pages from three different types of sources. 9 of them are health related government Facebook pages, 11 of them are mainstream media's official health Facebook pages and 18 sources are pages of sources certified by Health on Net Foundation (HONcode), a non-profit organization in official relations with the World Health Organization (WHO) that promotes transparent and reliable health information online. From these 38 reliable media's Facebook pages, we collected 33561 articles for 01/01/2015 to 4/02/2018. 
\subsubsection{Unreliable Media}
Dr. Melissa Zimdars, a communication and media expert, prepared a list of false, misleading, clickbaity, and satirical media outlets~\cite{melissa2016list, wikimelissa}. Similar lists are also maintained by Wikipedia~\cite{wikifakelist} and \url{informationisbeautiful.net}~\cite{info2016list}. We identified $6$ media outlets which primarily spread health-related misinformation and are present in these lists. Another source for identifying unreliable outlets is \textit{Snopes.com}, a popular hoax-debunking website that fact-checks news of different domains including health. We followed the health or medical hoaxes debunked by \textit{Snopes.com} and identified $14$ media outlets which sourced those hoaxes. In total, we identified $20$ unreliable outlets. Table \ref{tab:list_outlets} lists the Facebook page ids of all the reliable and unreliable outlets that have been used in this study.

\begin{table}[t!]
\resizebox{\linewidth}{!}{
\begin{tabular}{l|l}
\toprule
Reliable   &        \begin{tabular}{@{}c@{}} everydayhealth, 
 WebMD, 
 statnews, 
 AmericanHeart, 
 BBCLifestyleHealth, \\
 CBSHealth, 
 FoxNewsHealth, 
 WellNYT, 
 latimesscience, 
 tampabaytimeshealth, \\
 philly.comhealth, 
 AmericanHeart, 
 AmericanCancerSociety, 
 HHS, 
 CNNHealth, \\
 cancer.gov, 
 FDA, 
 mplus.gov, 
 NHLBI, 
 kidshealthparents, \\
 ahrq.gov, 
 healthadvocateinc, 
 HealthCentral, 
 eMedicineHealth, 
 C4YWH, \\
 BabyCenter, 
 MayoClinic, 
 MedicineNet, 
 healthline\end{tabular}\\
\midrule
Unreliable &        \begin{tabular}{@{}c@{}} liveahealth, 
 healthexpertgroup, 
 healthysolo, 
 organichealthcorner, \\
 justhealthylifestyle1, 
 REALfarmacyCOM, 
 thetruthaboutcancer, 
 BookforHealthyLife, \\
 viralstories.bm, 
 justhealthyway, 
 thereadersfile, 
 pinoyhomeremedies, \\
 onlygenuinehealth, 
 greatremediesgreathealth, 
 HealthRanger, 
 thefoodbabe, \\
 AgeofAutism, 
 HealthNutNews, 
 consciouslifenews, 
 HealthImpactNews\end{tabular}\\
\bottomrule
\end{tabular}%
}
\caption{List of Facebook page ids of the reliable and unreliable outlets. Some of them are unavailable now.}
\label{tab:list_outlets}
\vspace{-10mm}
\end{table}

% Without expertise in health field, it was a tough task to label a source as unreliable. We chose three different ways to get unreliable sources. Facebook pages of sources listed by Melisa Zimdars in her 'False, Misleading, Clickbait-y, and Satirical "News" Sources', pages of sources listed by Information is Beautiful and Facebook pages sharing the health news that were labeled as false in Snopes.com, a fact-checking website. 
\subsection{Data Collection}
The next challenge is to gather news articles published by the selected outlets. We identified the official Facebook pages of each of the $49$ media outlets and collected all the link-posts~\footnote{Facebook allows posting status, pictures, videos, events, links, etc. We collected the link type posts only.} shared by the outlets within January 1, 2015 and April 2, 2018~\footnote{After that, Facebook limited access to pages as a result of the Cambridge Analytica incident.} using Facebook Graph API. For each post, we gathered the corresponding news article link, the status message, and the posting date.

\subsubsection{News Article Scraping} We used a Python package named Newspaper3k~\footnote{https://newspaper.readthedocs.io/en/latest/} to gather the news article related data. Given a news article link, this package provides the headline, body, author name (if present), and publish date of the article. It also provides the visual elements (image, video) used in an article. In total, we collected data for $29,047$ articles from reliable outlets and $15,017$ from unreliable outlets.
% explain why FB is needed. We collected 21800 articles for 01/01/2015 to 4/02/2018. Despite the higher number of unreliable Facebook pages, we got fewer number of articles. The reasons for low articles are the sources being banned or not on service and sources are created after 2015. 
\subsubsection{Filtering non-Health News Articles}
\begin{figure}[h]
\vspace{-5mm}
\includegraphics[width=2in]{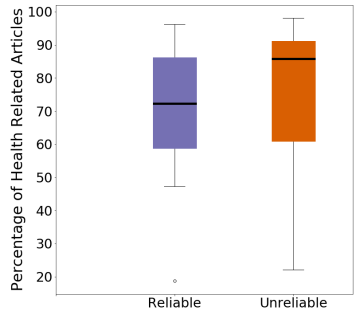}
\centering
\vspace{-5mm}
\caption{Comparison between reliable and unreliable outlets with respect to presence of health-related news contents}
\label{fig: boxplot_health_related_articles}
\vspace{-5mm}
\end{figure}

Even though we restricted ourselves to health-related outlets, we observed that the outlets also published or shared non-health (e.g., sports, entertainment, weather) news. We removed these non-health articles from our dataset and only kept \textit{health}, \textit{food \& drink}, or \textit{fitness \& beauty} related articles. Specifically, for each news article, we used the document categorization service provided by Google Cloud Natural Language API~\footnote{https://cloud.google.com/natural-language/} to determine its topic. If an article doesn't belong to one of the three above mentioned topics, it is filtered out. This step reduced the dataset size to $27,589$; $18,436$ from reliable outlets and $9,153$ from unreliable outlets. We used this health-related dataset only in all the experiments of this paper. Figure \ref{fig: boxplot_health_related_articles} shows the health-related news percentage distribution for reliable outlets and unreliable outlets using box-plots. For each of the $29$ reliable outlets, we measure the percentage of health news and then use these $29$ percentage values to draw the box-plot for the reliable outlets; likewise for unreliable. We observe that the reliable outlets (median 72\%) publish news on health topics comparatively less than unreliable outlets (median 85\%).

% , it is seen that the median for percentage of health-related articles i.e. articles with Google NLP category health, food \& drink, and beauty \& fitness for unreliable media is about 85 percent and about 72 percent for reliable media.

%One observation we made during this process is that unreliable outlets publish more non-health news articles compared to reliable outlets even though all these outlets are health related outlet.
%{Figure XX shows distributions of non-health/health article ratios for each outlet category. $50\%$ of the reliable outlets have a non-health/health ratio of XX whereas for the unreliable outlets, this number is YY.}

% \begin{figure}
% \includegraphics[width=\linewidth]{./fig/Top 10 Categories by percent in unreliable sources.eps}
% \centering
% \caption{ Top 10 Categories in unreliable sources}
% \label{fig: Top 10 Categories in unreliable sources}
% \end{figure}

% To filter out these non-health news articles, we   While we were on topic modeling of the articles, we notice that some of them were not related to health. To solve this problem, we categories the articles using the Natural Language API by Google cloud and filtered out non health related articles. We decided to work only on the articles with category `health', `food \& drink', and `beauty \& fitness'. We got 21105 out of 33561 articles from reliable sources to be health related. Whereas, only 9739 out of 21800 from unreliable sources were health relaated. One interesting observation we made after this filtration was unreliable sources doesn't share only one category of news. We theories that it is because they share anything that is viral.

% \subsubsection{Structure of the Data}
% \begin{itemize}
%     \item Status Message
%     \item Linkname/Headline
%     \item News Content
%     \item Links[]
%     \item Anchors []
%     \item Google Categories []
%     \item Author Name
%     \item Publish Date
%     \item Facebook Posting Date
%     \item Link of Visual Media
% \end{itemize}

\section{Analysis}
\label{sec-analysis}

Using this dataset, we conduct content analysis to examine structural, topical, and semantic differences in health news from reliable and unreliable outlets.
\subsection{Structural Difference} 
\subsubsection{Headline}
The headline is a key element of a news article.  According to a study done by American Press Institute and the Associated Press~\cite{headlineonly}, only $4$ out of $10$ Americans read beyond the headline. So, it is important to understand how reliable and unreliable outlets construct the headlines of their health-related news. According to to~\cite{chartbeat}, a long headline results in significantly higher click-through-rate (CTR) than a short headline does. We observe that the average headline length of an article from reliable outlets and an article from unreliable outlets is $8.56$ words and $12.13$ words, respectively. So, on average, an unreliable outlet's headline has a higher chance of receiving more clicks or attention than a reliable outlet's headline. To further investigate this, we examine the \textit{clickbaityness} of the headlines. The term clickbait refers to a form of web content (headline, image, thumbnail, etc.) that employs writing formulas, linguistic techniques, and suspense creating visual elements to trick readers into clicking links, but does not deliver on its promises~\cite{clickbaitdefinition}. Chen et al. \cite{chen2015misleading} reported that clickbait usage is a common pattern in false news articles. We investigate to what extent the reliable and unreliable outlets use clickbait headlines in their health articles. For each article headline, we test whether it is a clickbait or not using two supervised clickbait detection models-- a sub-word embedding based deep learning model ~\cite{rony2017diving} and a feature engineering based Multinomial Naive Bayes model \cite{clkbtdetectorsaurabh}. Agreement between these models was measured as $0.44$ using Cohen's $\kappa$. We mark a headline as a clickbait if both models labeled it as clickbait. We observe, 27.29\% (5,031 out of 18,436) of the headlines from reliable outlets are click bait. In unreliable outlets, the percentage is significantly higher, 40.03\% (3,664 out of 9,153). So, it is evident that the unreliable outlets use more click baits than reliable outlets in their health journalism. 
% This finding is consistent with a previous research \cite{rony2017diving} 

We further investigate the linguistic patterns used in the clickbait headlines. In particular, we analyze the presence of some common patterns which are generally employed in clickbait according to~\cite{chartbeat, outbrain}. The patterns are-
\begin{itemize}
    \item Presence of demonstrative adjectives (e.g., this, these, that)
    \item Presence of numbers (e.g., 10, ten)
    \item Presence of modal words (e.g., must, should, could, can)
    \item Presence of question or WH words (e.g., what, who, how)
    \item Presence of superlative words (e.g., best, worst, never) 
\end{itemize}
% Linguistic features are often helpful to identify misleading content. That's why we analyzed the headline of the articles and extracted different linguistic features. We used seven different features and identified whether a headline contains any of them. ~\ref{fig:linguistic_feature} shows the distribution of the features. It is clear from the analysis that unreliable media use the demonstrative adjective, number, modal, and w/h questions more in comparison with reliable media. But reliable media use more question mark in their headlines than the unreliable media do. 
\begin{figure}
\includegraphics[width=\linewidth]{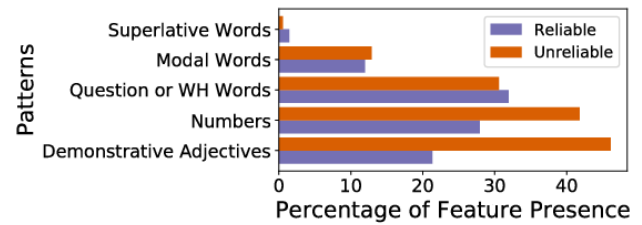}
\centering
\vspace{-5mm}
\caption{Distribution of clickbait patterns}
\label{fig:linguistic_feature}
\vspace{-5mm}
\end{figure}
Figure \ref{fig:linguistic_feature} shows the distribution of these patterns among the clickbait headlines of reliable and unreliable outlets. Note, one headline may contain more than one pattern. For example, this headline \textit{``Are these the worst 9 diseases in the world?''} contains four of the above patterns. This is the reason why summation of the percentages isn't equal to one.  We see that unreliable outlets use demonstrative adjective and numbers significantly more compared to the reliable outlets.

\subsubsection{Time-span Between Publishing and Sharing}
\begin{figure}[h]
\includegraphics[width=\linewidth]{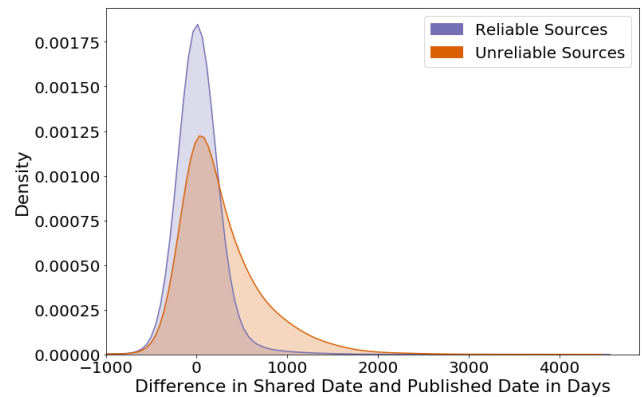}
\centering
\vspace{-5mm}
\caption{Distribution of (Shared Date - Published Date) gaps in days}
\label{fig: densityplot_Shared_difference}
\vspace{-4mm}
\end{figure}
We investigate the time difference between an article's published date and share date (in Facebook). Figure \ref{fig: densityplot_Shared_difference} shows density plots of \emph{Facebook Share Date -- Article Publish Date} for reliable and unreliable outlets. We observe that both outlet categories share their articles on Facebook within a short period after publishing. However, unreliable outlets seem to have considerable time gap compared to reliable outlets. It could be because of re-sharing an article after a long period. To verify that, we checked how often an article is re-shared on Facebook. We find that on average a reliable article is shared 1.057 times whereas an unreliable article is shared 1.222 times.

\begin{figure*}[!h]
    \centering
    \begin{subfigure}{0.33\textwidth}
        \centering
        \includegraphics[width=\linewidth]{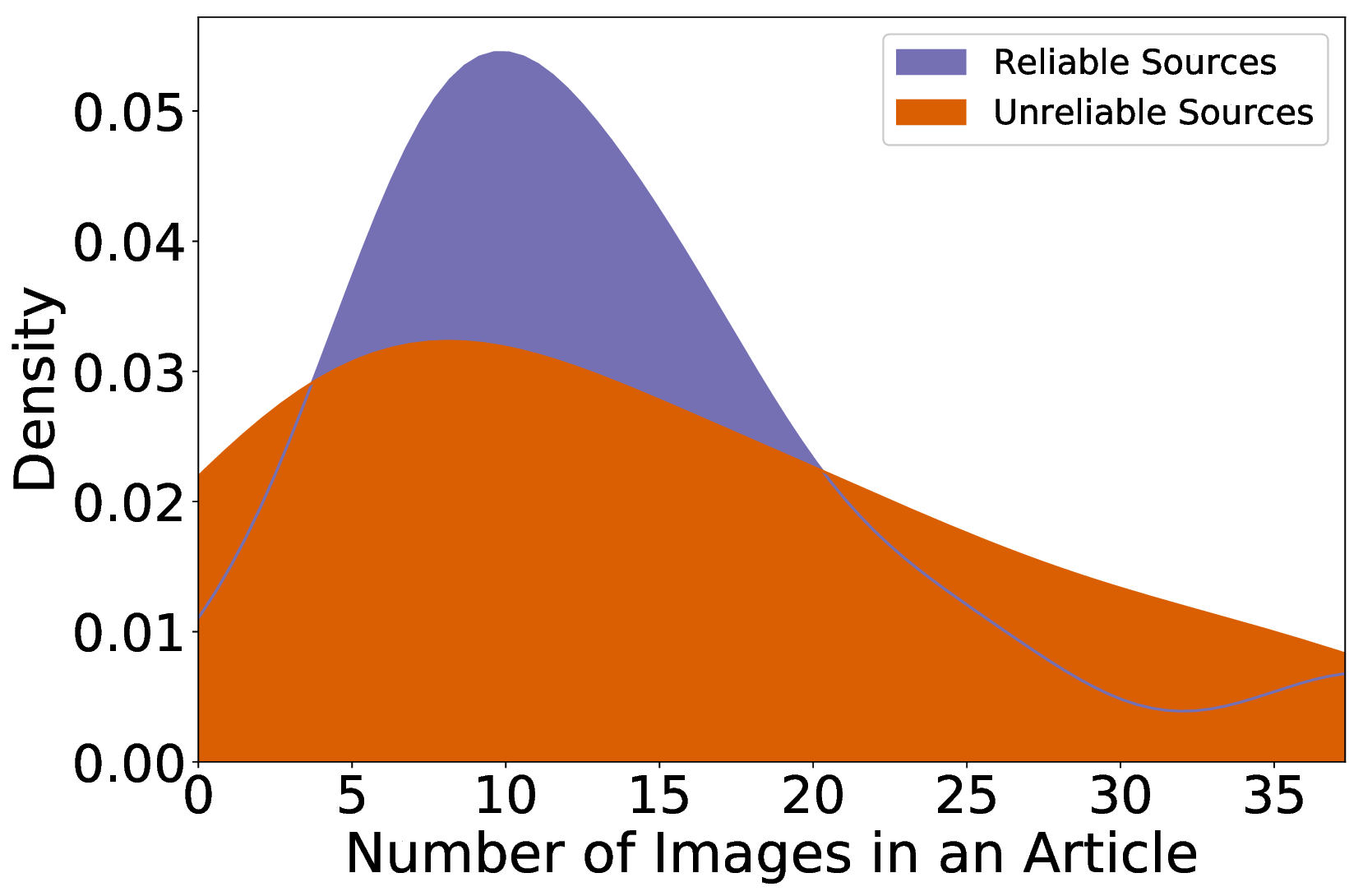}
        \caption{Image}
        \label{fig: densityplot_images}
    \end{subfigure}%
%     \hspace{2mm}
    \begin{subfigure}{0.33\linewidth}
        \centering
        \includegraphics[width=\linewidth]{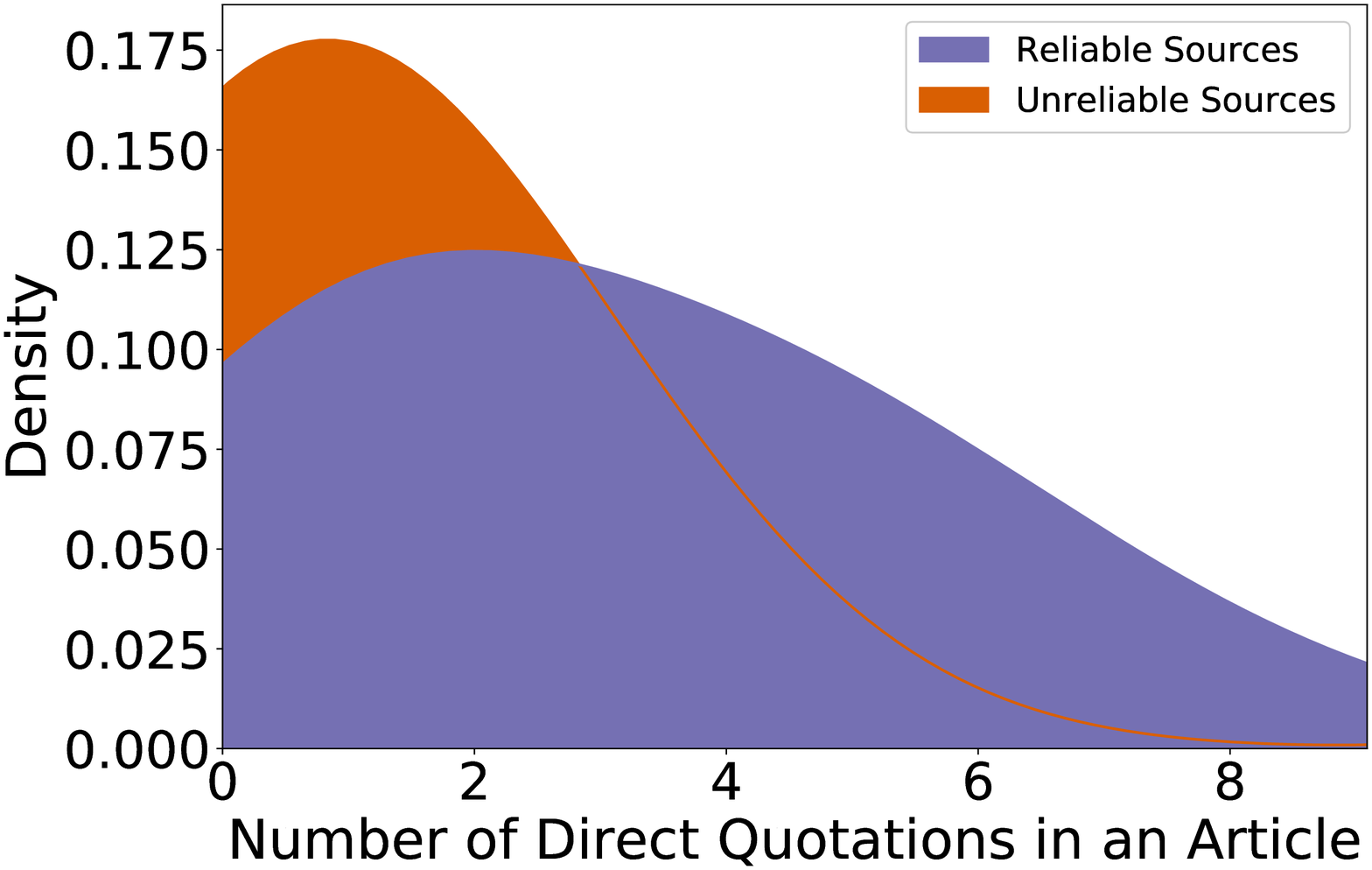}
        \caption{Quotation}
        \label{fig: densityplot_quotation}
    \end{subfigure}%
%     \hspace{2mm}
    \begin{subfigure}{0.33\linewidth}
        \centering
        \includegraphics[width=\linewidth]{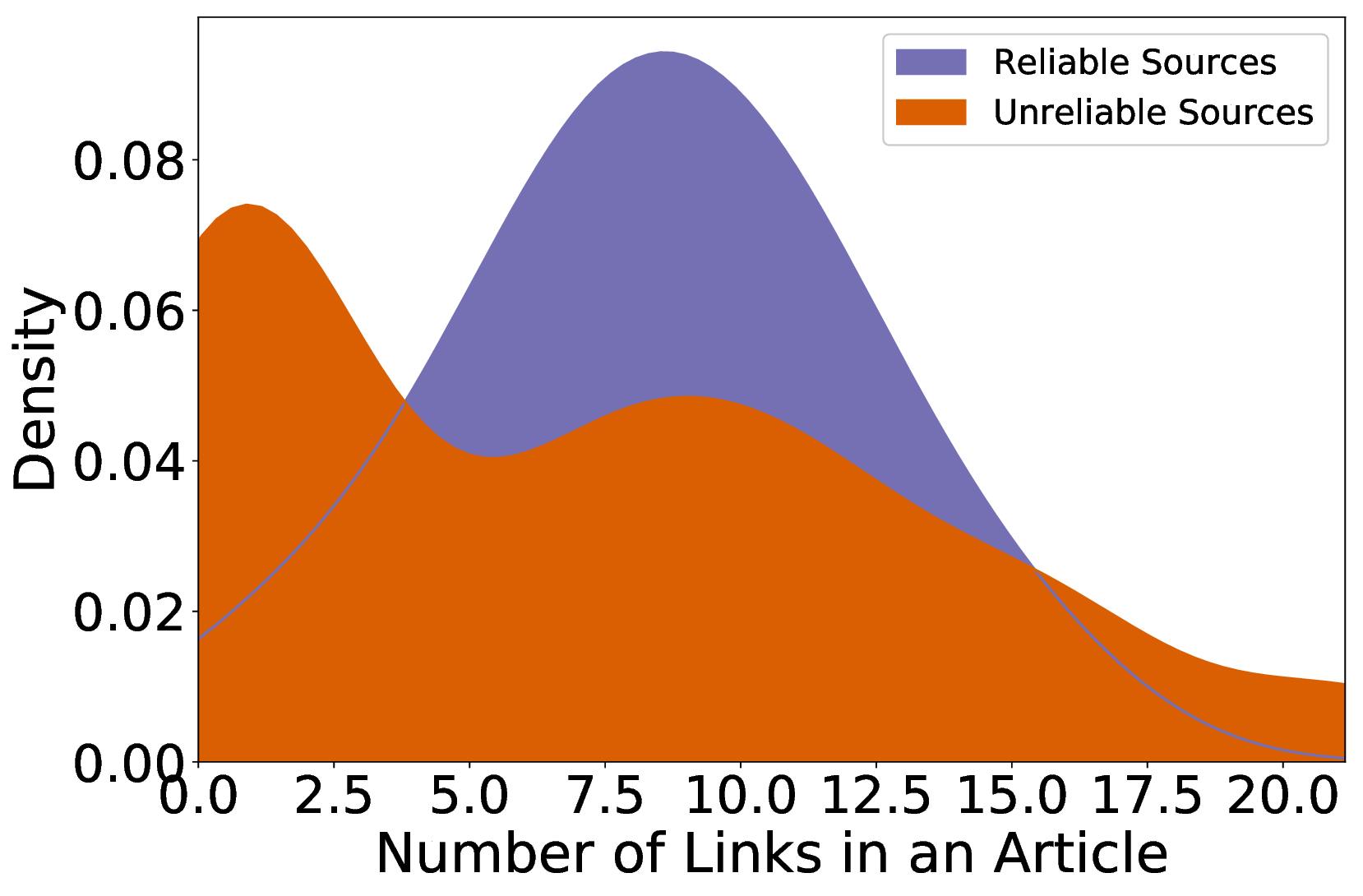}
        \caption{Link}
        \label{fig: densityplot_link}
    \end{subfigure}
    
    \label{fig:densities}
    \vspace{-4mm}
    \caption{Distribution of average number of image/quotation/link per article from reliable and unreliable outlets.}
\end{figure*}

\begin{figure}[!h]
    \centering
    \begin{subfigure}{0.33\linewidth}
        \centering
        \includegraphics[width=\linewidth]{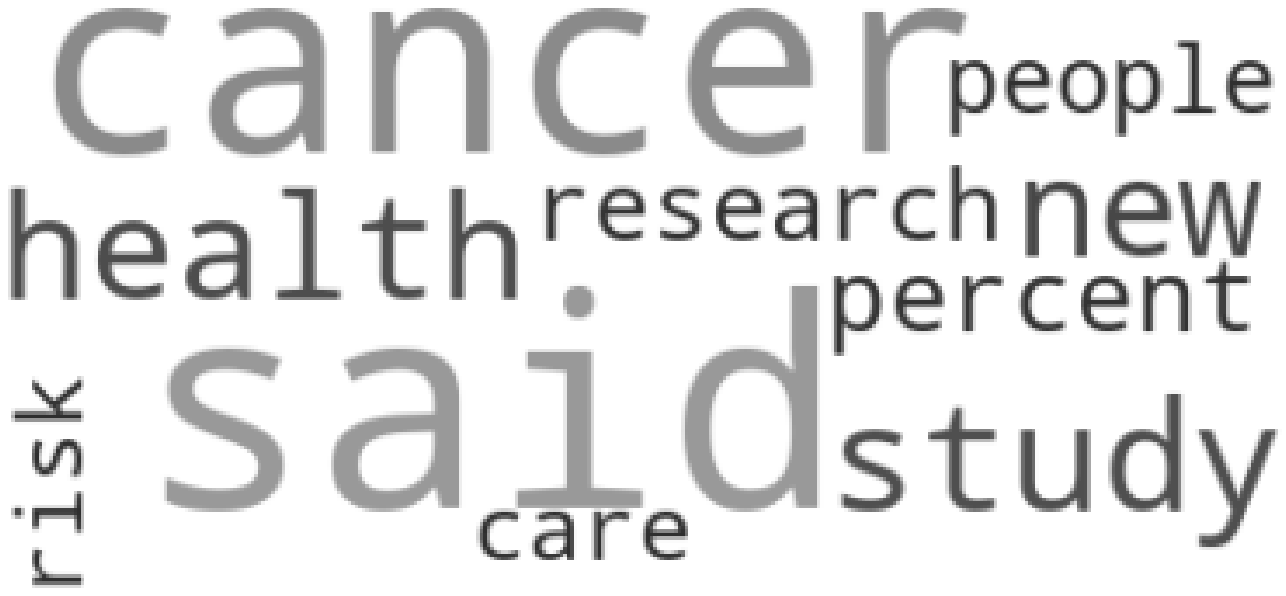}
        \caption{RT1}
        \label{fig:rt1}
    \end{subfigure}%
%     \hspace{2mm}
    \begin{subfigure}{0.33\linewidth}
        \centering
        \includegraphics[width=\linewidth]{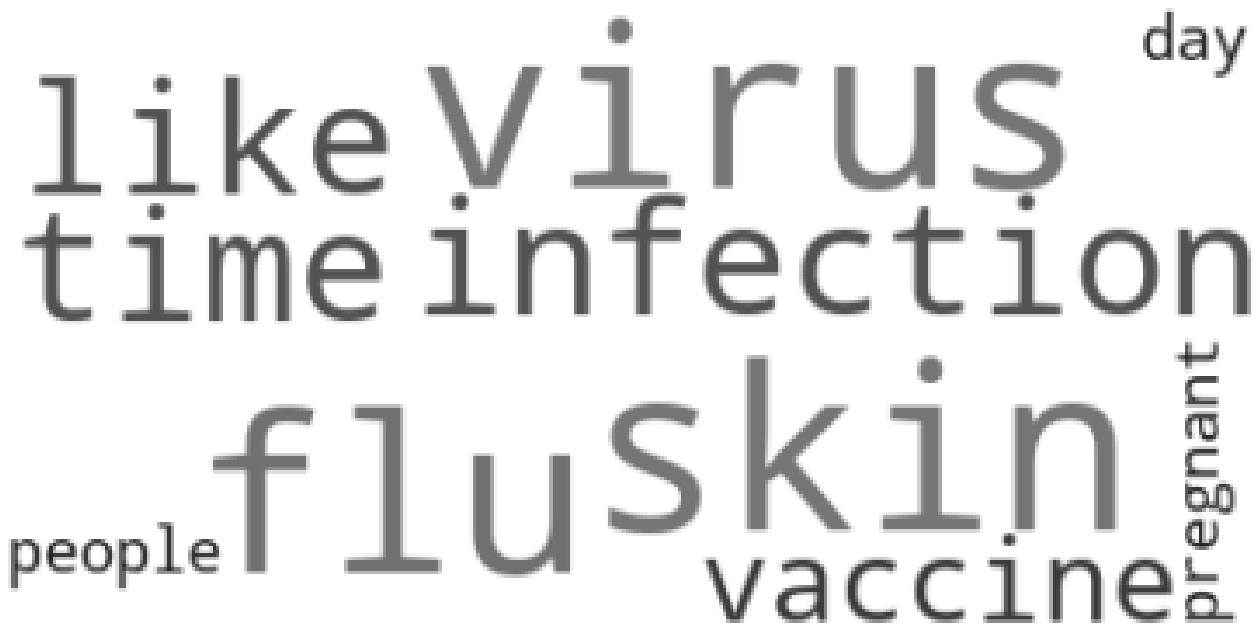}
        \caption{RT2}
        \label{fig:rt2}
    \end{subfigure}%
%     \hspace{2mm}
    \begin{subfigure}{0.33\linewidth}
        \centering
        \includegraphics[width=\linewidth]{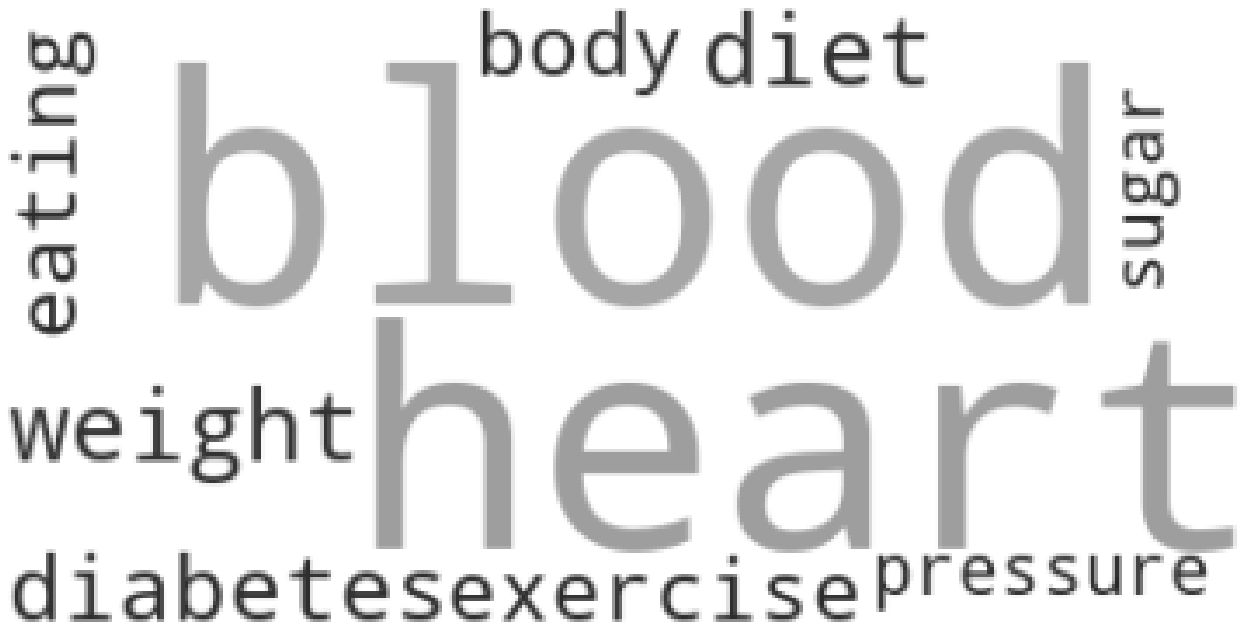}
        \caption{RT3}
        \label{fig:rt3}
    \end{subfigure}

    \begin{subfigure}{0.33\linewidth}
        \centering
        \includegraphics[width=\linewidth]{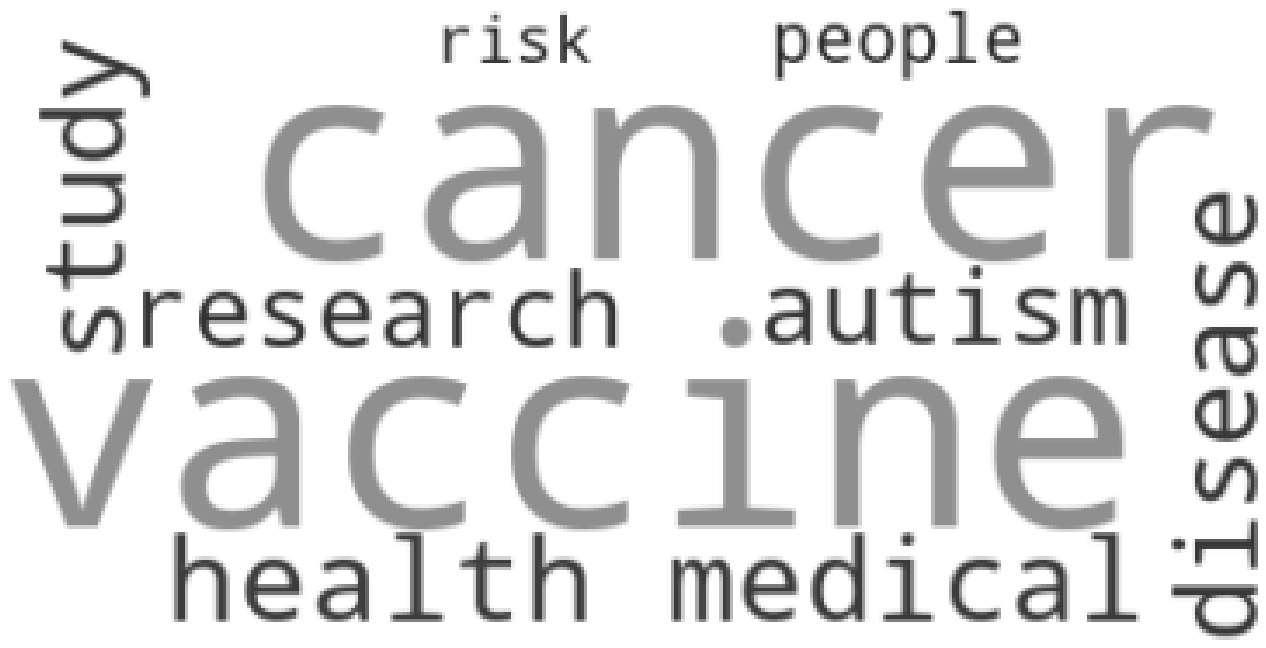}
        \caption{UT1}
        \label{fig:ut1}
    \end{subfigure}%
%     \hspace{2mm}
    \begin{subfigure}{0.33\linewidth}
        \centering
        \includegraphics[width=\linewidth]{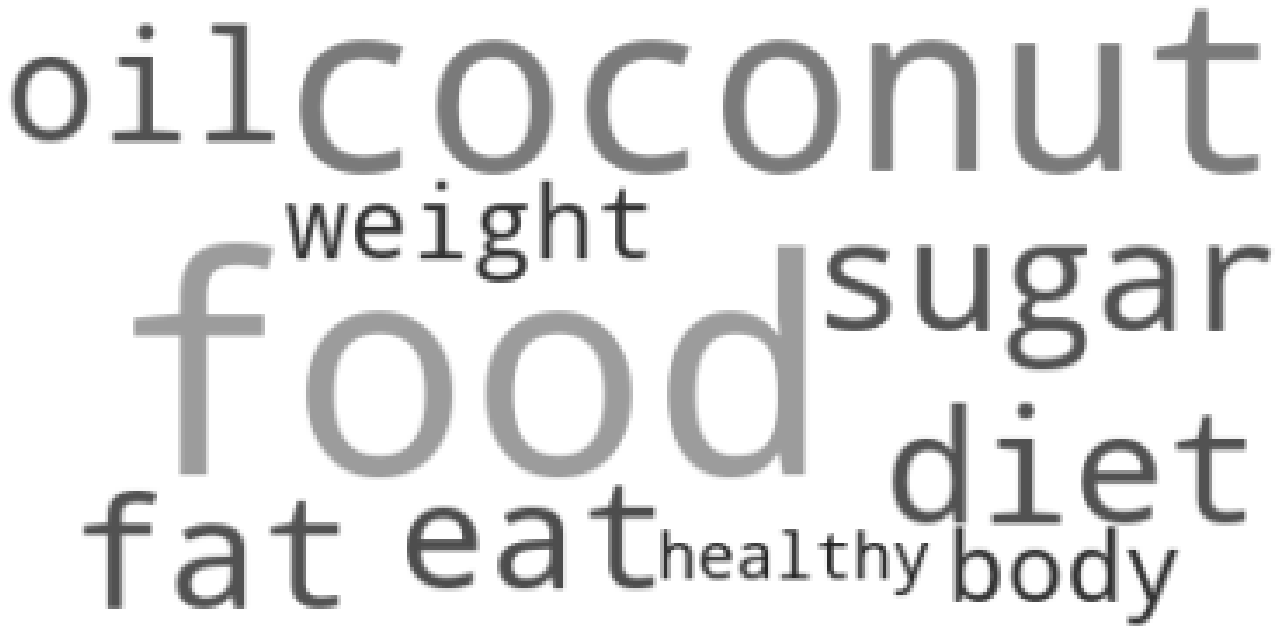}
        \caption{UT2}
        \label{fig:ut2}
    \end{subfigure}%
%     \hspace{2mm}
    \begin{subfigure}{0.33\linewidth}
        \centering
        \includegraphics[width=\linewidth]{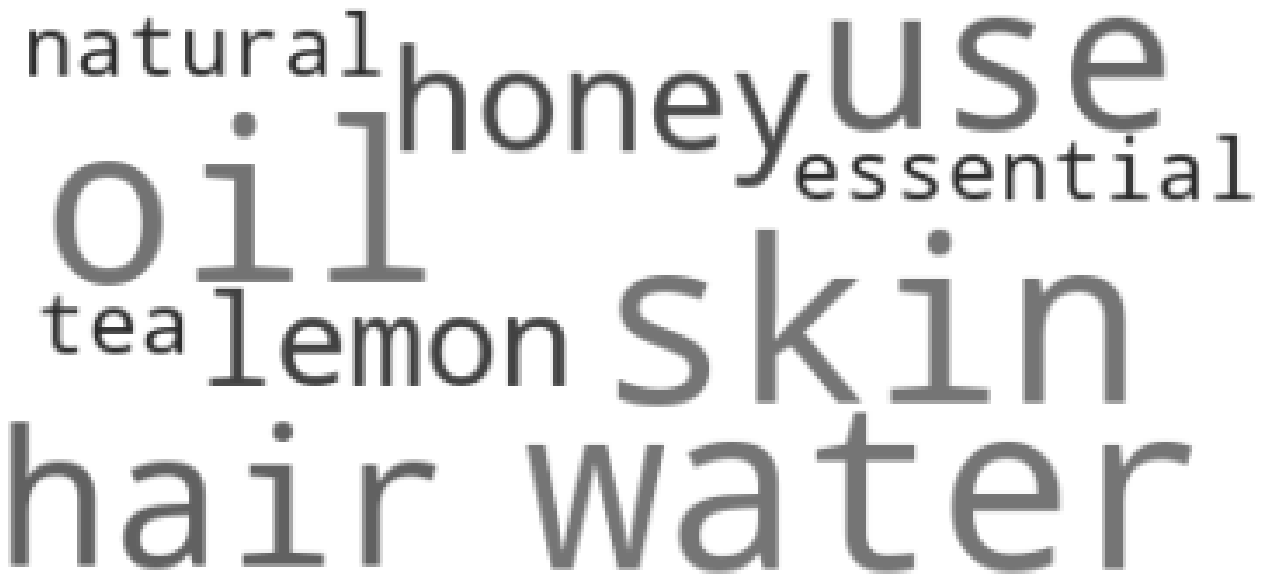}
        \caption{UT3}
        \label{fig:ut3}
    \end{subfigure}
    
    \vspace{-3mm}
    \caption{Topic modeling ($k=3$) of articles from reliable outlets (top, denoted as RT) and from unreliable outlets (bottom, denoted as UT).}
    \label{fig:topicmodeling}
    \vspace{-5mm}
\end{figure}

\subsubsection{Use of visual media}
We examined how often the outlets use images in the articles. Our analysis finds that on average an article from reliable outlets uses 13.83 images and an article from unreliable outlets uses 14.22 images. Figure \ref{fig: densityplot_images} shows density plots of the average number of images per article for both outlet categories. We observe that a good portion of unreliable outlet sourced articles uses a high number of images (more than 20).

\subsection{Topical Difference}
All the articles which we examined are health-related. However, the health domain is considerably broad and it covers many topics. We hypothesize that there are differences between the health topics which are discussed in reliable outlets and in unreliable outlets. To test that, we conduct an unsupervised and a supervised analysis.
\subsubsection{Topic Modeling}
We use \textit{Latent Dirichlet Allocation(LDA)} algorithm to model the topics in the news articles. The number of topics, $k$, was set as 3. Figure \ref{fig:topicmodeling} shows three topics for each of the outlet categories. Each topic is modeled by the top-10 important words in that topic. The font size of words is proportional to the importance. Figure \ref{fig:rt1} and \ref{fig:ut1} indicate that ``cancer'' is a common topic in reliable and unreliable outlets. Although, the words \emph{study}, \emph{said}, \emph{percent}, \emph{research}, and their font sizes in Figure \ref{fig:rt1} indicate that the topic ``cancer'' is associated with research studies, facts, and references in reliable outlets. On the contrary, unreliable outlets have the words \emph{vaccine}, \emph{autism}, and \emph{risk} in Figure \ref{fig:ut1} which suggests the discussion regarding how vaccines put people under autism and cancer risk, an unsubstantiated claim, generally propagated by unreliable media~\footnote{https://www.webmd.com/brain/autism/do-vaccines-cause-autism}$^,$\footnote{https://www.skepticalraptor.com/skepticalraptorblog.php/polio-vaccine-causes-cancer-myth/}. Figure \ref{fig:ut2} and \ref{fig:ut3} suggest the discussions about weight loss, skin, and hair care products (e.g., essential oil, lemon). Topics in Figure \ref{fig:rt2} and \ref{fig:rt3} discuss mostly  flu, virus, skin infection, exercise, diabetes and so on.

\begin{figure}[!h]
    \centering
    \begin{subfigure}{0.54\linewidth}
        \centering
        \includegraphics[width=\linewidth]{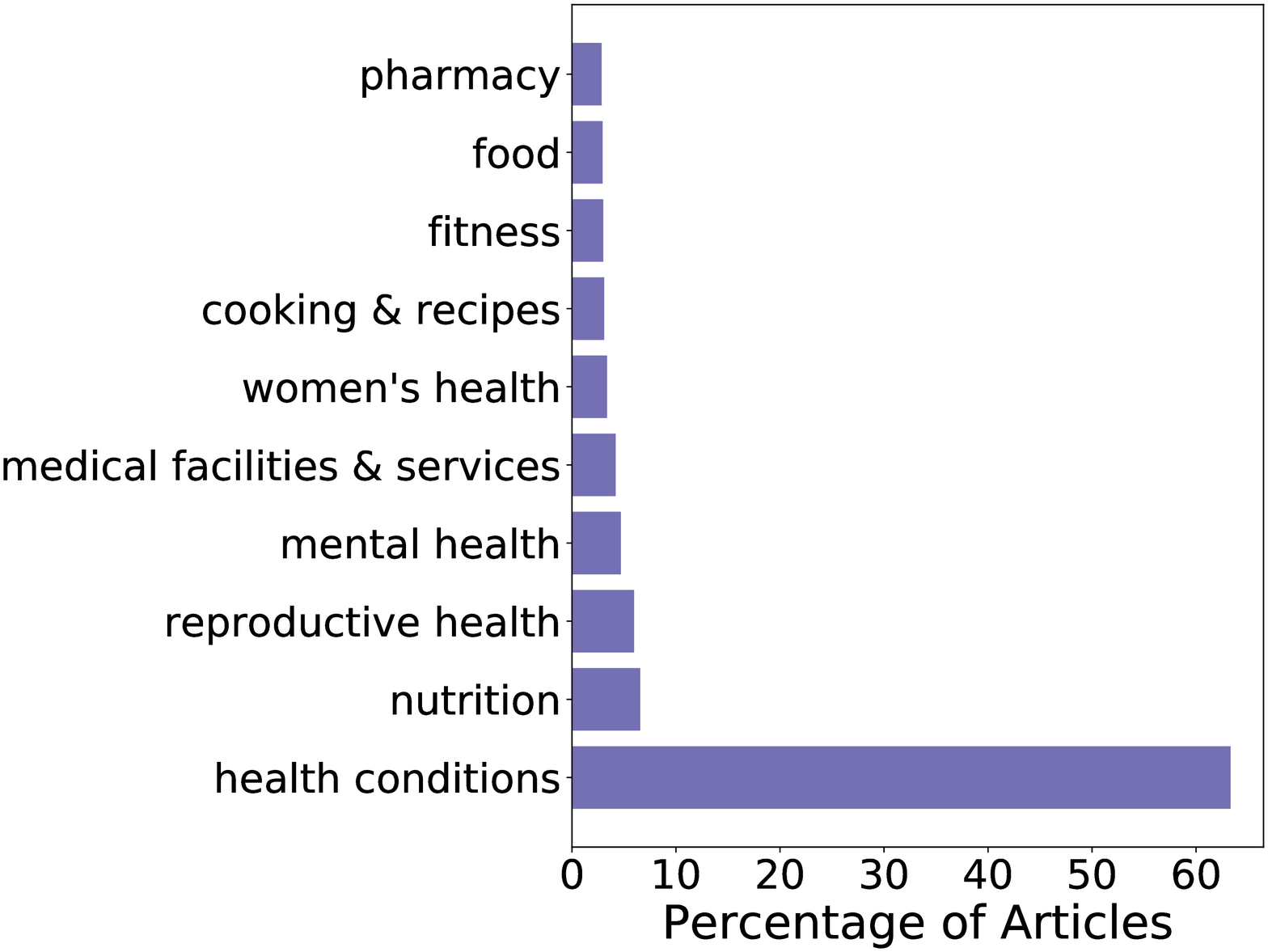}
        \caption{Reliable}
        \label{fig:reliable_sub-categories}
    \end{subfigure}%
%     \hspace{2mm}
    \begin{subfigure}{0.46\linewidth}
        \centering
        \includegraphics[width=\linewidth]{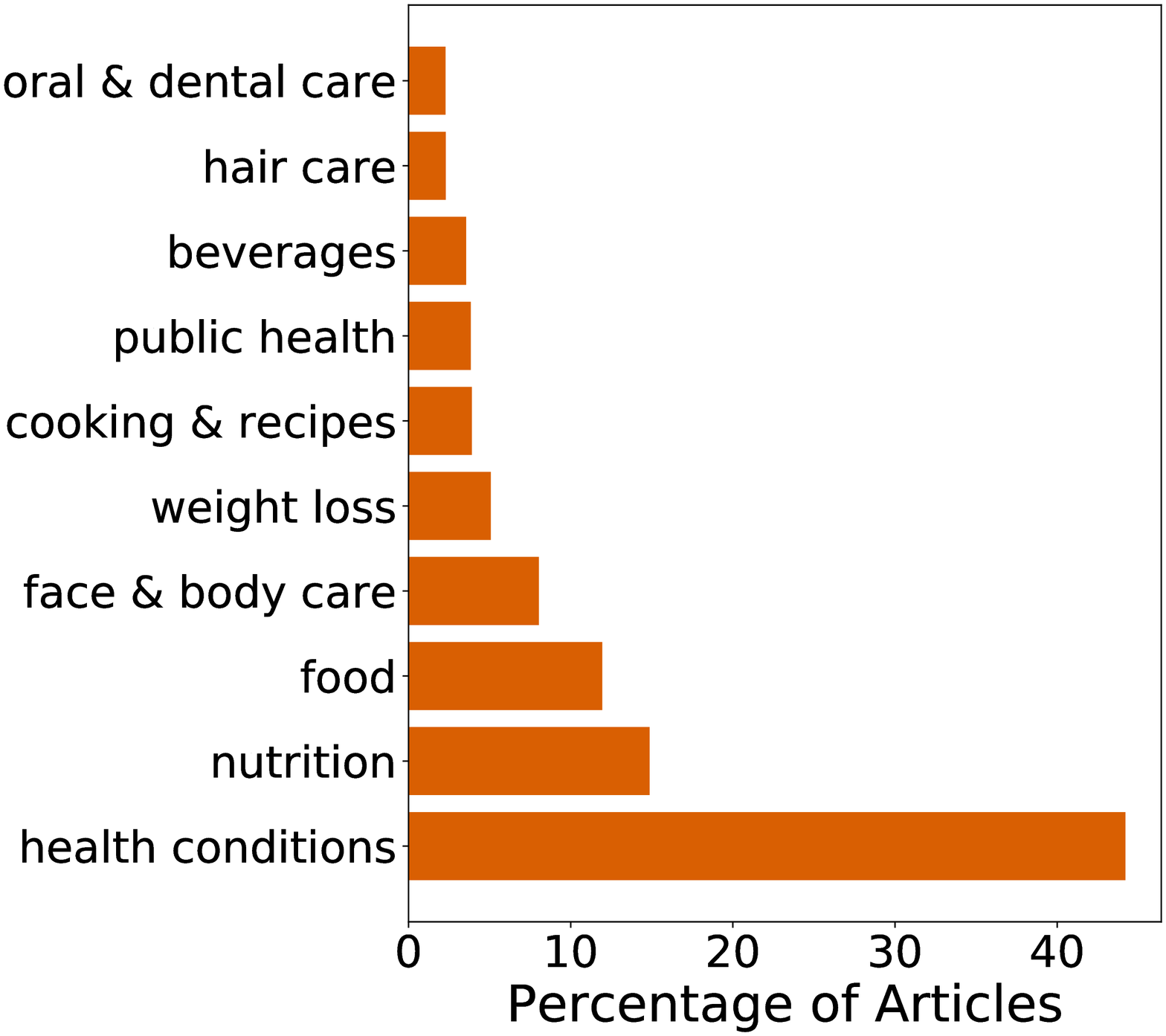}
        \caption{Unreliable}
        \label{fig:unreliable_sub-categories}
    \end{subfigure}%

    \caption{Top-10 topics in reliable and unreliable outlets.}
    \label{fig:subcategories}
\end{figure}

\subsubsection{Topic Categorization}
In addition to topic modeling, we categorically analyze the articles' topics using Google Cloud Natural Language API~\footnote{https://cloud.google.com/natural-language/}. Figure \ref{fig:subcategories} shows the top-10 topics in the reliable and unreliable outlets. In the case of reliable, the distribution is significantly dominated by \emph{health condition}. On the other hand, in the case of unreliable outlets, percentages of \emph{nutrition} and \emph{food} are noticeable. Only 4 of the 10 categories are common in two outlet groups. Unreliable topics have \emph{weight loss}, \emph{hair care}, \emph{face \& body care}. This finding supports our claim from topic modeling analysis.

% %sub categories figures for Reliable sources
% \begin{figure}[h]
% \includegraphics[width=50mm,scale=0.5]{./fig/reliable_sub-categories.eps}
% \centering
% \caption{Top Ten Sub-categories for Reliable media}
% \label{fig:reliable_sub-categories.eps}
% \end{figure}

% %sub categories figures for Unreliable sources
% \begin{figure}[h]
% \includegraphics[width=50mm,scale=0.5]{./fig/unreliable_sub-categories.eps}
% \centering
% \caption{Top Ten Sub-categories under Health for Unreliable media}
% \label{fig:unreliable_sub-categories}
% \end{figure}

% Figure \reminder{\ref{fig: Top_10_Categories_by_percent_in_Unreliable_Sources}} and figure \reminder{\ref{fig: Top_10_Categories_by_percent_in_Reliable_Sources}} show that articles with category 'health' is more in reliable media than in unreliable media. Unreliable media are using nutriment and beauty to engage with their audiences compare to reliable media. 

\subsection{Semantic Difference}
We analyze what efforts the outlets make to make a logical and meaningful health news. Specifically, we consider to what extent the outlets use quotations and hyperlinks. Use of quotation and hyperlinks in a news article is associated with credibility \cite{sundar1998effect,de2012journalistic}. Presence of quotation and hyperlinks indicates that an article is logically constructed and supported with credible factual information. 

\subsubsection{Quotation} We use the Stanford QuoteAnnotator~\footnote{https://stanfordnlp.github.io/CoreNLP/quote.html} to identify the quotations from a news article. Figure \ref{fig: densityplot_quotation} shows density plots of the number of quotations per article for reliable and unreliable outlets. We observe that unreliable outlets use less number of quotations compared to reliable outlets. We find that the average number of quotations per article is 1 and 3 in unreliable and reliable outlets, respectively. This suggests that the reliable outlet sources articles are more credible and unreliable outlets are less credible.
% \begin{figure}[h]
% \includegraphics[width=\linewidth]{./fig/densityplot_avg_total_quotations.eps}
% \centering
% \caption{Average number of total Quotations}
% \label{fig: densityplot_total_quotations}
% \end{figure}

% \begin{figure}[h]
% \includegraphics[width=\linewidth]{./fig/densityplot_avg_direct_quotations.eps}
% \centering
% \caption{Average number of Direct Quotations}
% \label{fig: densityplot_direct_quotations}
% \end{figure}

% Average number of total quotations for reliable media is 4.51 quotes in an article and for unreliable sources it is 2.85 quotes per article. 
% figure \reminder{\ref{fig: densityplot_total_quotations}} shows the distribution. Average number of direct quotations for reliable media is 3 quotes with the maximum of 9.09 in an article and for unreliable sources it is 1 quote per article with the maximum of 3.34.  
% figure \reminder{\ref{fig: densityplot_total_quotations}} shows the distribution. 

% Reliable media includes more quotations using different experts as to back up the articles. From topic modeling we saw "said" was dominating. Unreliable media lack the quotation, keeping us wondering if the points they make are just opinions. From the topic modeling we also observed that they used word evidence to try to prove their point. If the point is fact there should be a source too.  

%Convert /cancer to www.example/cancer. Is it true that unreliable outlets use inhouse links more often than reliable outlets.

\subsubsection{Hyperlink}
% \begin{figure}[h]
% \includegraphics[width=\linewidth]{./fig/densityplot_median_links.eps}
% \centering
% \caption{Average links}
% \label{fig: densityplot_median_link}
% \end{figure}
We examine the use of the hyperlink in the articles. On average, a reliable outlet sourced article contains 8.4 hyperlinks and an unreliable outlet sourced article contains 6.8 hyperlinks. Figure \ref{fig: densityplot_link} shows density plots of the number of links per article for reliable and unreliable outlets. The peaks indicate that most of the articles from reliable outlets have close to 8 (median) hyperlinks. On the other hand, most of the unreliable outlet articles have less than 2 hyperlinks. This analysis again suggests that the reliable sourced articles are more credible than unreliable outlet articles.
% Another difference between articles in reliable and unreliable sources is the number of links present in an article. The articles in reliable sources have average of 8.4 links and unreliable sources have 6.8 links. Figure \reminder{\ref{fig: densityplot_median_link}}  shows the distribution of median of number of links in reliable and unreliable sources. Most of the sources in reliable media have medians around 7 whereas medians of sources in unreliable media are around 0. 

% \subsubsection{Factual Statements}

\section{Conclusion and Future Work}
\label{sec-discussion}
In this paper, we closely looked at structural, topical, and semantic differences between articles from reliable and unreliable outlets. Our findings reconfirm some of the existing claims such as unreliable outlets use clickbaity headlines to catch the attention of users. In addition, this study finds new patterns that can potentially help separate health disinformation. For example, we find that less quotation and hyperlinks are more associated with unreliable outlets. However, there are some limitations to this study. For instance, we didn't consider the videos, cited experts, comments of the users, and other information. In the future, we want to overcome these limitations and leverage the findings of this study to combat health disinformation.
% \begin{itemize}
%     \item A fake health news detector. Our analysis identifies possible features.
%     \item Limitation. Didn't consider ads. Didn't consider print media.
% \end{itemize}
% \input{sec-notes}

\bibliographystyle{ACM-Reference-Format}
\bibliography{reference}

\end{document}